\documentstyle[12pt,epsf]{article}
\def\singlespace {\smallskipamount=3.75pt plus1pt minus1pt
                  \medskipamount=7.5pt plus2pt minus2pt
                  \bigskipamount=15pt plus4pt minus4pt
                  \normalbaselineskip=15pt plus0pt minus0pt
                  \normallineskip=1pt
                  \normallineskiplimit=0pt
                  \jot=3.75pt
                  {\def\smallskip {\vskip\smallskipamount}}
                  {\def\medskip   {\vskip\medskipamount}}
                  {\def\bigskip   {\vskip\bigskipamount}}
                  {\setbox\strutbox=\hbox{\vrule
                    height10.5pt depth4.5pt width 0pt}}
                  \parskip 7.5pt
                  \normalbaselines}
\def\middlespace {\smallskipamount=5.825pt plus1.5pt minus1.5pt
                  \medskipamount=11.25pt plus3pt minus3pt
                  \bigskipamount=22.5pt plus6pt minus6pt
                  \normalbaselineskip=22.5pt plus0pt minus0pt
                  \normallineskip=1pt
                  \normallineskiplimit=0pt
                  \jot=5.825pt
                  {\def\smallskip {\vskip\smallskipamount}}
                  {\def\medskip   {\vskip\medskipamount}}
                  {\def\bigskip   {\vskip\bigskipamount}}
                  {\setbox\strutbox=\hbox{\vrule
                    height15.75pt depth6.75pt width 0pt}}
                  \parskip 7.25pt
                  \normalbaselines}
\def\dblspc {\smallskipamount=7.5pt plus2pt minus2pt
                  \medskipamount=15pt plus4pt minus4pt
                  \bigskipamount=30pt plus8pt minus8pt
                  \normalbaselineskip=30pt plus0pt minus0pt
                  \normallineskip=2pt
                  \normallineskiplimit=0pt
                  \jot=7.5pt
                  {\def\smallskip {\vskip\smallskipamount}}
                  {\def\medskip   {\vskip\medskipamount}}
                  {\def\bigskip   {\vskip\bigskipamount}}
                  {\setbox\strutbox=\hbox{\vrule
                    height21.0pt depth9.0pt width 0pt}}
                  \parskip 15.0pt
                  \normalbaselines}

\def\pr{\prime }

\def\be{\begin{equation}}
\def\lan{\left\langle}
\def\ran{\right\rangle}

\def\j-{\J_-}

\def\ee{\end{equation}}

\def\bearr{\begin{eqnarray}}
\def\bearrs{\begin{eqnarray*}}
\def\eearr{\end{eqnarray}}
\def\eearrs{\end{eqnarray*}}
\def\barr{\begin{array}}
\def\earr{\end{array}}

\def\nn8{\nonumber\\[10pt]}
\def\l{\left}
\def\r{\right}

\def\ve{\varepsilon}

\def\dis{\displaystyle}

\def\ed{\end{document}}

\begin{document}
\singlespace

\begin{center}
{\bf Transition Strength Sums and Quantum Chaos in Shell Model States} 
\vskip 0.25cm

V.K.B. Kota$^{a,}$ {\footnote{Invited talk in the {\it National
Seminars on Nuclear Physics} held at Institute of Physics,
Bhubaneswar during July 26-29,
1999.}}, R. Sahu$^{b}$, K. Kar$^{c}$, J.M.G. G\'omez$^{d}$
and J. Retamosa$^{d}$  \\
$^{a}${\it Physical Research Laboratory, Ahmedabad \,\,380 009, India} \\
$^{b}${\it Physics Department, Berhampur University,
Berhampur\,\, 760 007, India} \\
$^{c}${\it Saha Institute of Nuclear Physics, 1/AF
Bidhannagar, \\ Calcutta 700 064, India} \\
$^{d}${\it Departamento de F$\acute{i}$sica At$\acute{o}$mica, 
Molecular y Nuclear, Universidad Complutense de Madrid, E-28040 Madrid, Spain}
\end{center}
\vskip 0.5cm

{\small
\noindent {\bf ABSTRACT:} For the embedded Gaussian orthogonal
ensemble (EGOE) of random matrices, the strength sums generated
by a transition operator acting on an eigenstate  vary with the
excitation energy as the ratio of two Gaussians. This general
result is compared to exact shell model calculations, with
realistic interactions, of spherical orbit occupancies and
Gamow-Teller strength sums in some $(ds)$ and $(fp)$ shell
examples.  In order to confirm that EGOE operates in the chaotic
domain of the shell model spectrum, calculations are carried out
using two different interpolating hamiltonians generating
order-chaos transitions.  Good agreement is obtained in the
chaotic domain of the spectrum, and strong deviations are
observed as nuclear motion approaches a regular regime
(transition strength sums appear to follow the Dyson's
$\Delta_3$ statistic).  More importantly, they shed new light on
the newly emerging understanding that in the chaotic domain of
isolated finite interacting many particle systems smoothed
densities (they include strength functions) define the
statistical description of these systems and these densities
follow from embedded random matrix ensembles; some EGOE calculations
to this end are presented.

\begin{center}
{\bf 1. Introduction}
\end{center}

\noindent In the last fifteen years there has been an explosive
growth in the use of random matrix theories for quantum systems
particularly in the context of quntum chaos \cite{Gh-98,Br-81}.
Recently, working with the aim of developing a statistical
theory for finite interacting many particle systems, such as
atoms, molecules, nuclei, atomic clusters, metallic quantum dots
etc., by incorporating the ideas of random matrices and chaos,
several research groups recognized the importance of
investigating the embedded random matrix ensembles in detail,
i.e. the embedded Gaussian orthogonal ensemble of random
matrices of $k$-body interactions (EGOE($k$)) and their various
deformations [2-7]. The EGOE is introduced in the context of
nuclear shell model studies \cite{Br-81}; The EGOE($k$) is defined in
$m$-particle spaces (i.e. in the {\scriptsize $\l(\barr{c} {\cal N} \\ m
\earr \r)$} dimensional space generated by distributing
$m$ fermions over ${\cal N}$ single particle states) with a GOE
representation in $k$-particle space for $k$-body operators
(usually $k << m$). This ensemble and its deformed versions are
studied using Monte-Carlo methods with EGOE constructed in
occupation number representation and alternatively assuming
that realistic nuclear (similarly atomic) shell model hamiltonians are typical
members of EGOE(2) and its deformations, generic features of
these ensembles are inferred from exact shell model calculations.
For finite
interacting particle systems in the chaotic domain there are now
several studies of the statistical properties (both smoothed
forms and fluctuations) of energy levels, wavefunction
amplitudes or equivalently transition strengths generated by
action of a transition operator on an eigenstate and strength
functions [1-11].  However, only in the last 2-3 years similar
studies (i.e. in the context of quantum chaos) on expectation
values of operators, which measure transition strength sums, as
function of excitation energy have began [3,8,12-14].  
Given an operator $K={\cal O}^\dagger
{\cal O}$, the expectation values $\lan K
\ran^E$ are the diagonal elements of $K$ in the hamiltonian
($H$) diagonal basis (a more precise definition is given ahead
in Eq. (3)); they give strength sums generated by the transition
operator ${\cal O}$ acting on the eigenstate with energy $E$.
Two major examples are single particle transfer strength sums
which are expectation values of number operators that give
occupancies of single particle states (they determine
thermodynamic behaviour) and Gamow-Teller (GT) strength
sums as function of excitation energy which are relevant in
astrophysics (presupernova evolution and stellar collapse).  It
is expected that the smoothed $\lan K \ran^E$ vs $E$ will give
information about order-chaos transitions just as energies,
wavefunction amplitudes and transition strengths. 

Two important results given by EGOE are that in strongly
interacting shell model spaces (essentially in $0\hbar\omega$
spaces) (i) the state densities take Gaussian form
\cite{Mo-75} and (ii) the bivariate strength densities
take bivariate Gaussian form \cite{Fr-88}.  These results have
their basis in the EGOE representation of the hamiltonian $H$
(which is in general one plus two-body in nuclear case) and
transition operators ${\cal O}$.   The eigenvalue density $I(E)$
or its normalized version $\rho(E)$ is defined by \cite{Mo-75}
\be
\barr{c}
I(E) =  \langle\langle \delta (H-E) \rangle\rangle
= d\;\langle \delta (H-E) \rangle = d\; \rho(E)\;\;;\nn8
\rho(E)  \stackrel{{\mbox{EGOE}}}\longrightarrow {\overline{\rho(E)}} =
\rho_{\cal G}(E) = \dis\frac{1}{\dis\sqrt{2\pi} \sigma} exp -
\dis\frac{1}{2} \l(\dis\frac{E-\epsilon}{\sigma}\r)^2
\earr
\ee
In (1) $\lan\lan \cdots \ran\ran$ denotes trace (similary $\lan
\cdots \ran$ denotes average), the $\epsilon$, $\sigma$ and $d$
are centroid, width ($\sigma^2$ is variance) and dimensionality
respectively. Note that $\epsilon = \lan H \ran$, $\sigma^2=
\lan (H-\epsilon)^2 \ran$, `${\cal G}$' stands for Gaussian and
the bar over $\rho(E)$ indicates ensemble average (smoothed)
with respect to EGOE.  The strength $R(E,E^\pr )$ generated by a
transition operator ${\cal O}$ in the $H$-diagonal basis is
$R(E, E^\pr )$ = $\mid \lan E^\pr \mid {\cal O} \mid E \ran
\mid^2$.  Correspondingly the bivariate strength density
$I_{biv;{\cal O}}(E,E^\pr)$ or $\rho_{biv;{\cal O}}(E,E^\pr )$
which is positive definite and normalized to unity is \cite{Fr-88}
\be
\barr{c}
\barr{rcl}
I_{biv;{\cal O}}(E,E^\pr) & = &
\lan\lan {\cal O}^{\dag} \delta (H - E^\pr ) {\cal O}
\delta (H-E) \ran\ran \nn8
& = & I^\pr (E^\pr ) \mid \lan
E^\pr \mid {\cal O} \mid E \ran \mid^2 I(E) \nn8
& = & \lan\lan {\cal O}^{\dag} {\cal O} \ran\ran\;
\rho_{biv;{\cal O}}(E,E^\pr)\;;
\nn8
\earr
\nn8
\barr{c}
\rho_{biv;{\cal O}}(E,E^\pr) \stackrel{{\mbox{EGOE}}}\longrightarrow 
{\overline{\rho_{biv;{\cal O}}(E,E^\pr)}} = \rho_{biv-{\cal
G}; {\cal O}}(E,E^\pr) 
\earr
\earr
\ee
The bivariate Gaussian $\rho_{biv-{\cal G}; {\cal O}}(E,E^\pr) $
in (2) is defined by the centroids and variances of its marginal
densities 
and a bivariate correlation coefficient.  Though the EGOE forms
in (1,2) are derived by evaluating the averages over fixed-$m$
spaces, in large number of numerical shell model examples it is
verified that [2,10,15,16] they apply equally well in fixed-
$m$, $mT$ and $mJT$ spaces. In practice the so-called Edgeworth corrections
are added to the Gaussian forms in (1,2). Before
proceeding further two remarks are in order: 
\begin{enumerate}
\item Level and
strength fluctuations for EGOE follow \cite{Br-81} the GOE fluctuations, i.e.
nearest spacing distribution is Wigner distribution, Dyson-Mehta
$\Delta_3(L)$ statistic follow the GOE $ln(L)$ behaviour for
large $L$ and strength fluctuations are of Porter-Thomas (P-T)
type. The $\Delta_3$ form is tested ahead in Fig. 3 and recent
tests of P-T form for EGOE are given in Fig. 1.  

\item EGOE smoothed forms (1,2) (and (3) ahead) gave birth to the so
called statistical nuclear spectroscopy (see
[9,10,15-17] and references therein) and there are recent
studies of this in atoms \cite{Fl-94}, molecules and solids
\cite{Ka-94} and mesoscopic systems [3-5].
\end{enumerate}

The pupose of this paper is to first point out that EGOE, via
(1,2) gives rise to a statistical theory for the smoothed forms for
transition strength sums and the theory operates in the chaotic domain of the
spectrum. Shell model tests in $(ds)^6$ space for occupancies
and GT strength sums in $(fp)^6$ space are
carried out. The EGOE theory and the shell model studies are
described in Sect. 2. In order to confirm that the agreement
between shell model and EGOE theory is a consequence of
chaoticity of the shell model spectrum, GT strength sums and
occupancies are calculated using two different interpolating
hamiltonian that generate order-chaos transitions. Results of
these calculations form Sect. 3. Results in Sects. 2, 3 give
nuclear physics examples for the newly emerging understanding
that in the chaotic domain of isolated finite interacting many
particle systems smoothed densities, defined by EGOE, give rise to the
statistical description of these systems. Further remarks on
this important generic result for quantum chaos in isolated finite
interacting particle systems are given (together with the
results from a EGOE calculation for occupancies) in the concluding Section 4.

\begin{center}
{\bf 2. EGOE theory for transition strength sums and shell model tests}
\end{center}

One important by product of (2) is that the transition strength sum
density $\lan\lan {\cal O}^\dagger {\cal O} \delta(H-E)
\ran\ran$, which is a marginal density of the bivariate strength
density, takes a Gaussian form as the marginal of a bivariate
Gaussian is a Gaussian. Therefore, using (1), it is immediately
seen that transition strength sums vary with excitation energy
as ratio of Gaussians. Given $K={\cal O}^\dagger {\cal O}$, the
transition strength sum density is the expectation value density
denoted by $\rho_K(E)$ and then \cite{Kk-89,Fr-89},
\be
\barr{c}
\lan K\ran^E = [d(m)\rho(E)]^{-1}\;\l[\dis\sum_{\alpha \ve E} 
\lan E\alpha \mid K \mid E \alpha \ran \r] =
I_K(E)/I(E) = \rho_K(E)/\rho (E) \nn8
\stackrel{{\mbox{EGOE}}}\longrightarrow 
{\overline{\rho_K(E)}}/{\overline{\rho (E)}} = \rho_{K:{\cal G}}(E) / 
\rho_{\cal G}(E) \;\;,\nn8
\rho_K(E) = \lan K\delta (H-E)\ran
=d^{-1}\;I_K(E) = d^{-1}\;\lan\lan K \delta(H-E)\ran\ran \;;\;
K = {\cal O}^\dagger {\cal O}\;.
\earr
\ee
\newpage
\begin{center}
\epsfxsize 4.5in
\epsfysize 3in
\epsfbox{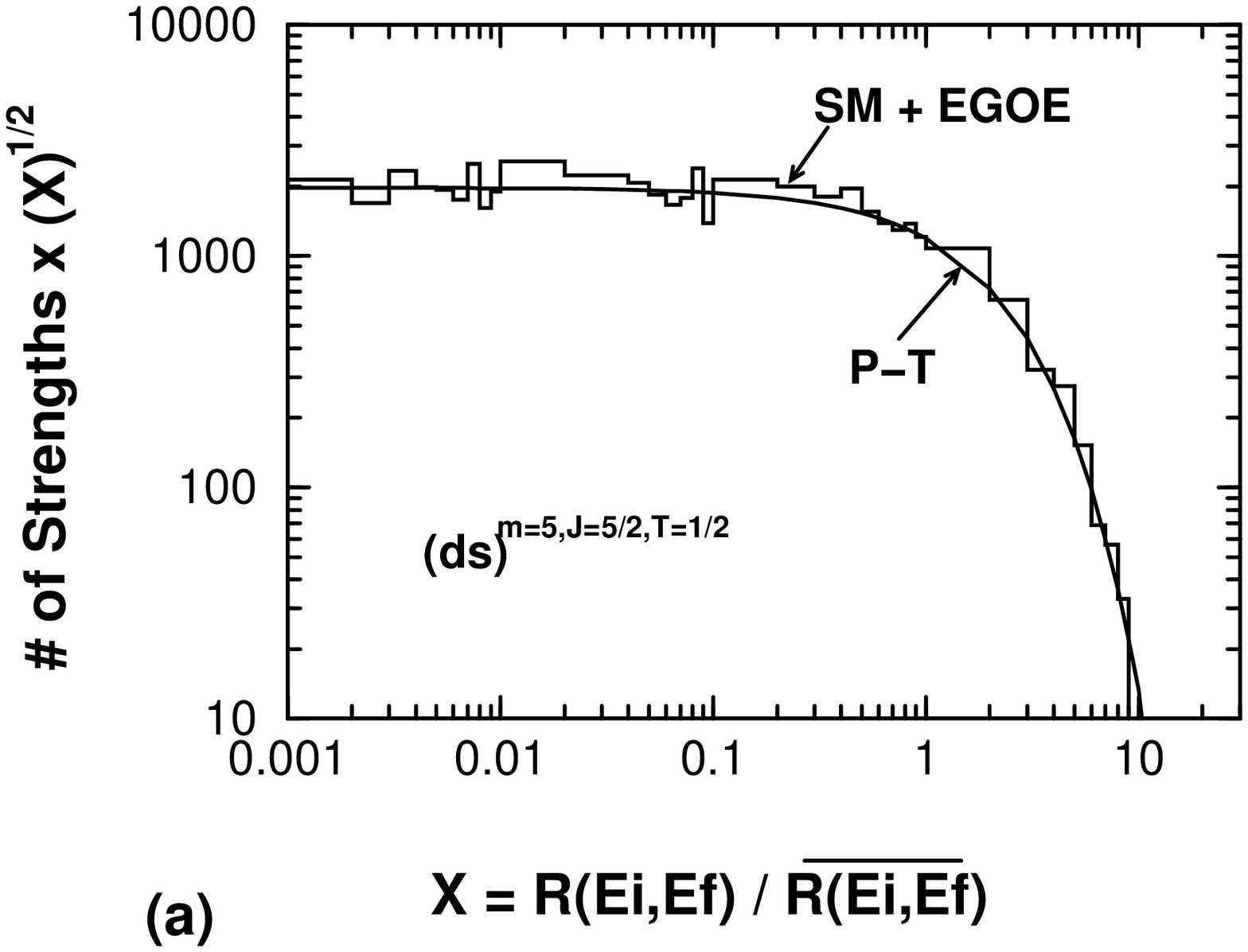}
\vskip -.75cm
\begin{tabular}{ll}
\epsfxsize 2.5in
\epsfysize 3.5in
\epsfbox{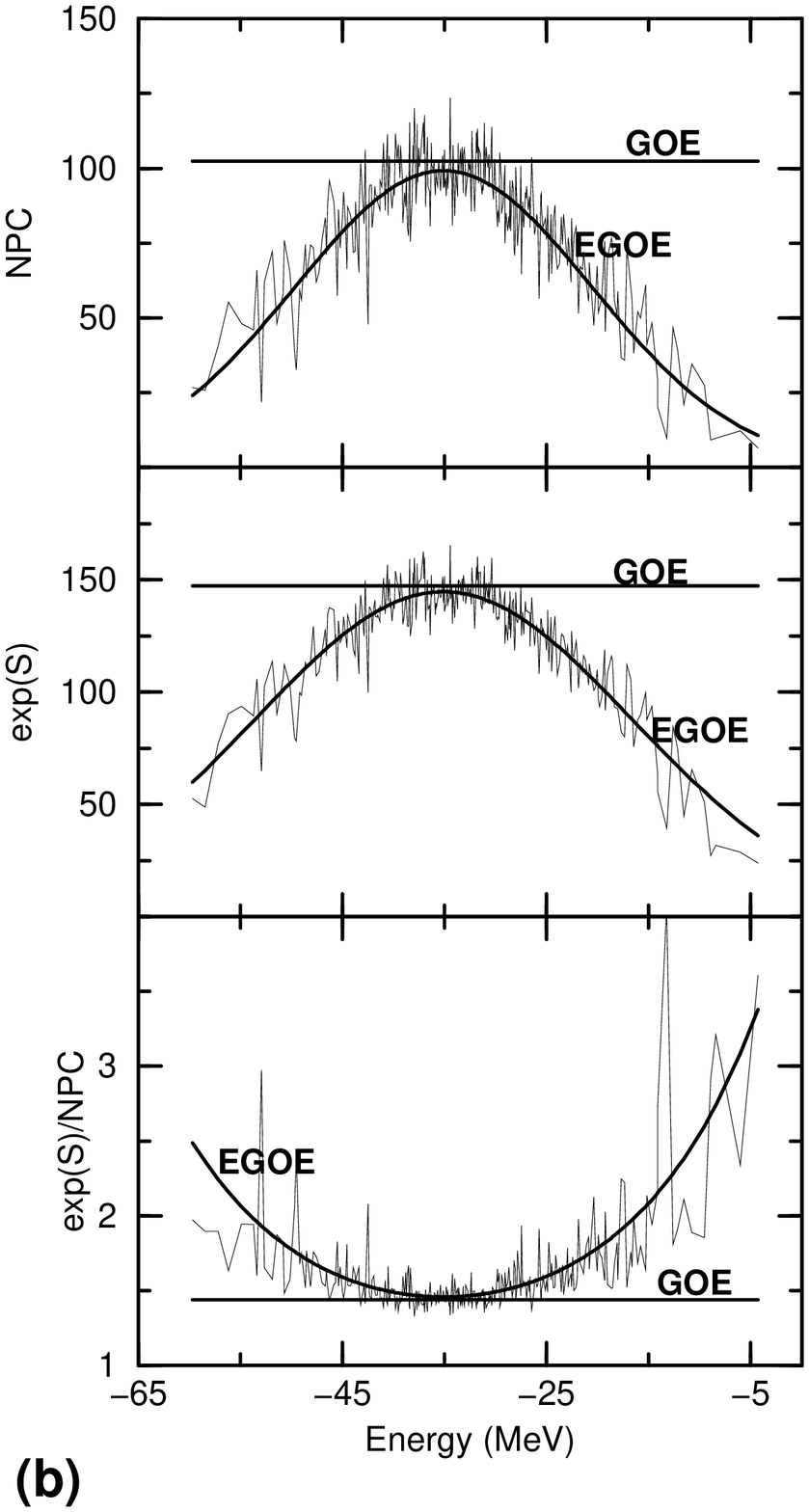}
&
\epsfxsize 2.5in
\epsfysize 3.5in
\epsfbox{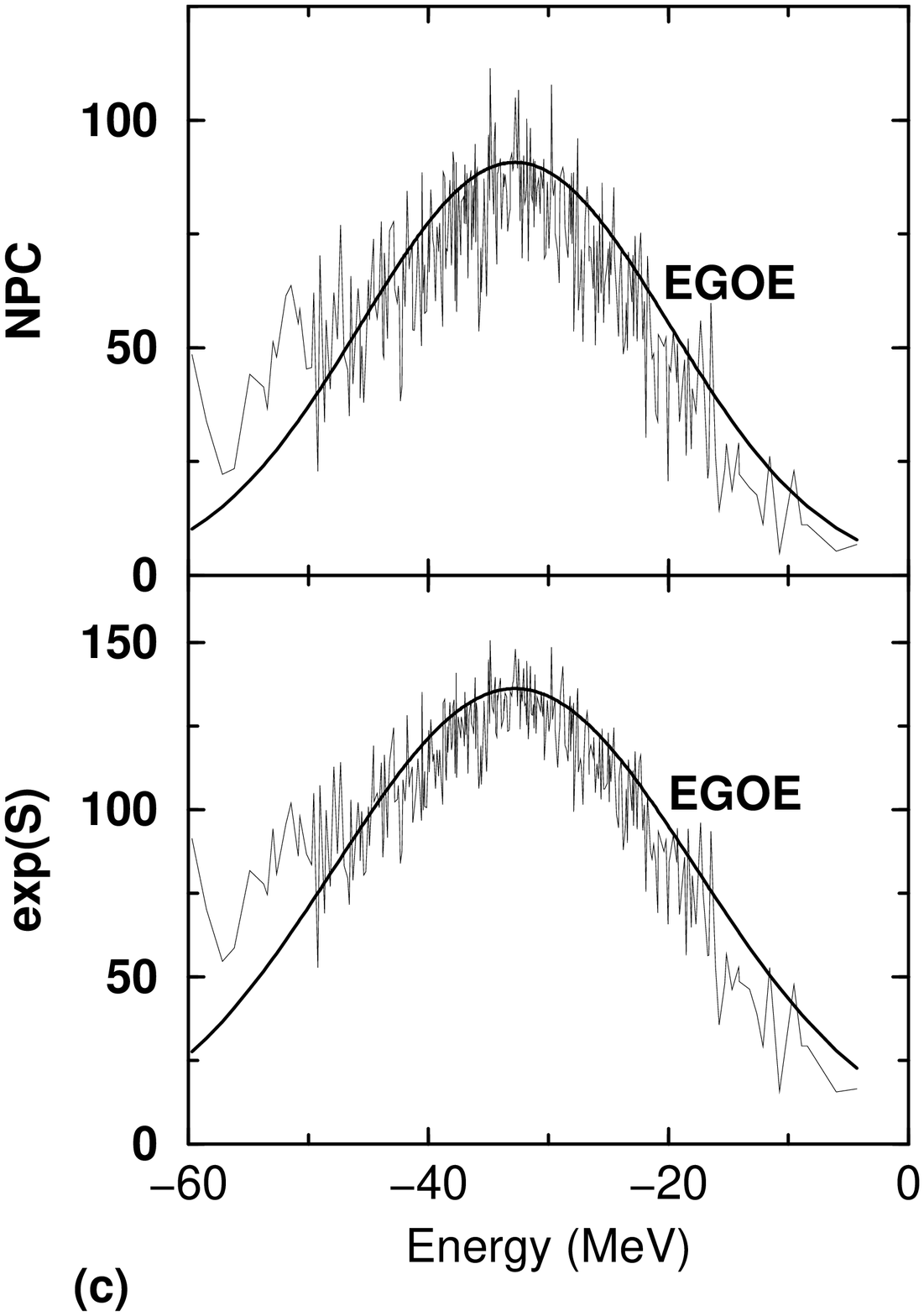}
\end{tabular}
\end{center}
\vskip -1cm
\noindent {\bf Fig. 1.} {\small {\bf (a)} Distribution of renormalized shell
model transition strengths  compared with the Porter-Thomas
(P-T) form. The locally averaged strengths
${\overline{R(Ei,Ef)}}$ are calculated using the EGOE
bivariate Gaussian form (2). Shell model (SM) calculations are
in 223 dimensional $(ds)^{m=5,J=5/2,T=1/2}$ space with hamiltonian defined by
Wildenthal interaction \protect\cite{Ze-96} and the transition
operator is the two-body part of the hamiltonian obtained after
substracting the configuration isospin centroid part.  The
SM+EGOE result is in good agreement with P-T. {\bf (b)} Number of
principal components (NPC) and information entropy ($S^{info}$)
versus energy ($E$) for a strength distribution in 307
dimensional $(ds)^{m=6,J=2,T=0}$ space. The hamiltonian is Kuo
interaction with $^{17}$O single particle energies
\protect\cite{Ku-67} and the transition operator is same as in (a) but 
for the Kuo interaction. The exact shell model
results are compared with the GOE and EGOE predictions; the EGOE
formulas (they use P-T) are given in \protect\cite{Ko-98}.
{\bf (c)} Same as (b) but for wavefunctions.}

\newpage
\noindent In deriving (3) it is assumed that the smoothed form of
$\rho_K(E)/\rho (E)$ reduces to ratio of smoothed forms of
$\rho_K(E)$ and $\rho (E)$.  This result ignores the
fluctuations in both $\rho_K(E)$ and $\rho (E)$ and the r.m.s
error due to neglect of the fluctuations is given in terms of
the number of principal components (NPC) or the inverse
participation ratio for the transition operator ${\cal O}$
\cite{Np-98}.  Note that (3) takes into account $(K,H)$ and $\l(
K,H^2 \r)$ correlations which define the centroid $\epsilon_K$
and width $\sigma_K$ of $\rho_K(E)$; $\epsilon_K=<KH>/<K>$ and
$\sigma_K^2=<KH^2>/<K> -\epsilon_K^2$. First discussions of the
EGOE result in (3) are in \cite{Kk-89,Ko-95,Fr-89}. Statistical
models that are inappropriate, are applied recently in the study
of GT strength sums as function of excitation enegy in
$(ds)$-shell \cite{Fr-97} although there are several studies of
GT strengths and strength sums in statistical nuclear
spectroscopy \cite{Kk-89,Ka-xx,Ks-89}.

In order to study the
domain of validity of (3), for the occupancies $\lan n_\alpha \ran^E$
shell model calculations in 307 dimensional $(2s 1d)^{m=6, J=2,
T=0}$ space are carried out using the Rochester - Oak Ridge
shell model code and for the GT strength sums in 814 
dimensional $(1f 2p)^{m=6, J=0,
T=0}$ space using the NATHAN code \cite{Ca-95} of the
Strasbourg-Madrid group. In the $(ds)$-shell studies the
hamiltonian employed is $H=h(1)+V(2)$ defined by Kuo's
\cite{Ku-67} two-body matrix elements ($V(2)$) and $^{17}$O
single particle energies ($h(1) \Longleftrightarrow
\epsilon_{d_{5/2}}=-4.15$ MeV, $\epsilon_{d_{3/2}}=0.93$ MeV,
$\epsilon_{s_{1/2}}=-3.28$ MeV).  Similarly in the $(fp)$ shell
studies the so called KB3 interaction \cite{Ca-95} is employed
with $h(1) \Longleftrightarrow \epsilon_{f_{7/2}}=0.0$ MeV,
$\epsilon_{f_{5/2}}=6.5$ MeV, $\epsilon_{p_{3/2}}=2$ MeV,
$\epsilon_{p_{1/2}}=4$ MeV).
The expectation value density $\rho_{n_\alpha :{\cal G}}$ for
the number operators $n_\alpha$ in $(ds)$ shell and
$\rho_{K(GT):{\cal G}}$ for the $K(GT)$ operator that generates
GT strength sums in $(fp)$ shell
are constructed in terms of their centriods and widths and
similarly the state density Gaussian. Then using (3) the
smoothed form of GT strength sum as function of excitation
energy is constructed and compared with exact shell model
results. From Fig. 2 it is seen that the EGOE result (3)
describes very well the shell model results except at the edges
of the spectra. The reason for the deviation at the edges is
well known - here the states are not chaotic (sufficiently
complex).  In the $(ds)$ shell example the $K$-density centroid
$\epsilon_K$, width $\sigma_K$, skewness $\gamma_{1:K}$ and
excess $\gamma_{2:K}$ (the $\gamma_1$ and $\gamma_2$ are shape
parameters that measure deviations from the Gaussian form) are
$\epsilon_K=-35.08$MeV, $\sigma_K=9.63$MeV, $\gamma_{1:K}=0.08$
and $\gamma_{2:K}=-0.09$ for $d_{5/2}$ density and ($-29.50$ MeV,
10.24 MeV, $-0.08$, $-0.17$) respectively for the $d_{3/2}$ density.
Similarly for the state density the parameters are
$\epsilon=-32.78$MeV, $\sigma=10.24$MeV, $\gamma_1=0.05$ and
$\gamma_2=-0.18$.  In the $(fp)$ shell example,
$\epsilon_K=11.12$MeV, $\sigma_K =8.65$MeV, $\gamma_{1:K}=0.09$,
$\gamma_{2:K}=-0.18$, $\epsilon=9.51$MeV, $\sigma=8.62$MeV,
$\gamma_1=0.10$ and $\gamma_2=-0.19$. Firstly the $|\gamma_1|$
and $|\gamma_2|$ values (being much less than $0.3$) clearly
show that all the densities are close to Gaussian.  Moreover
${\hat{\sigma}}= \sigma_K/\sigma \sim 1$. The scaled centroid
shifts ${\hat{\Delta}}_K=(\epsilon_K - \epsilon)/\sigma$ are
$-0.225$ and 0.32 for the $d_{5/2}$ and $d_{3/2}$ densities while
it is 0.187 for the $(fp)$ shell GT example. With $\lan K \ran^E$
given as ratio of Gausians and that ${\hat{\sigma}} \sim 1$
imply that in the middle of the spectrum $\lan K \ran^E \sim
\lan K \ran \l\{1+ {\hat{\Delta}}_K {\hat{E}} \r\}$;
${\hat{E}}=(E-\epsilon)/\sigma$, i.e $\lan K
\ran^E$ is linear in energy. This linear form and its polynomial
extensions are used in the past \cite{Ka-xx,Ks-89,Dr-77}.  The
value of ${\hat{E}}$ for the ground state are $-2.6$ and $-3.1$ for
$(ds)$ and $(fp)$ shell examples respectively. The range of
validity (from the center of the spectrum) of the linear form
for $\lan K \ran^E$ is much smaller for $(ds)$-shell than in
$(fp)$-shell as ${\hat{\Delta}}_K$ for the former is larger than
for the later; see \cite{Ks-89}.  Fig. 2 showing good agreement
between shell model results for $\lan n_\alpha \ran^E$  and
$\lan K({\mbox{GT}}) \ran^E$ with
realistic interactions and the EGOE prediction (3), gives a
justification to our assertion that EGOE starts operating from
the region of the onset of chaos for the strength sums as well.
Next we study this question via $\lan n_\alpha \ran^E$ and
$\lan K({\mbox{GT}}) \ran^E$ when the
hamiltonian changes thorough a parameter from a symmetry
preserving, i.e. regular hamiltonian to a chaotic one.
\vskip 1cm
\begin{center}
\begin{tabular}{ll}
\epsfxsize 2.5in
\epsfysize 6in
\epsfbox{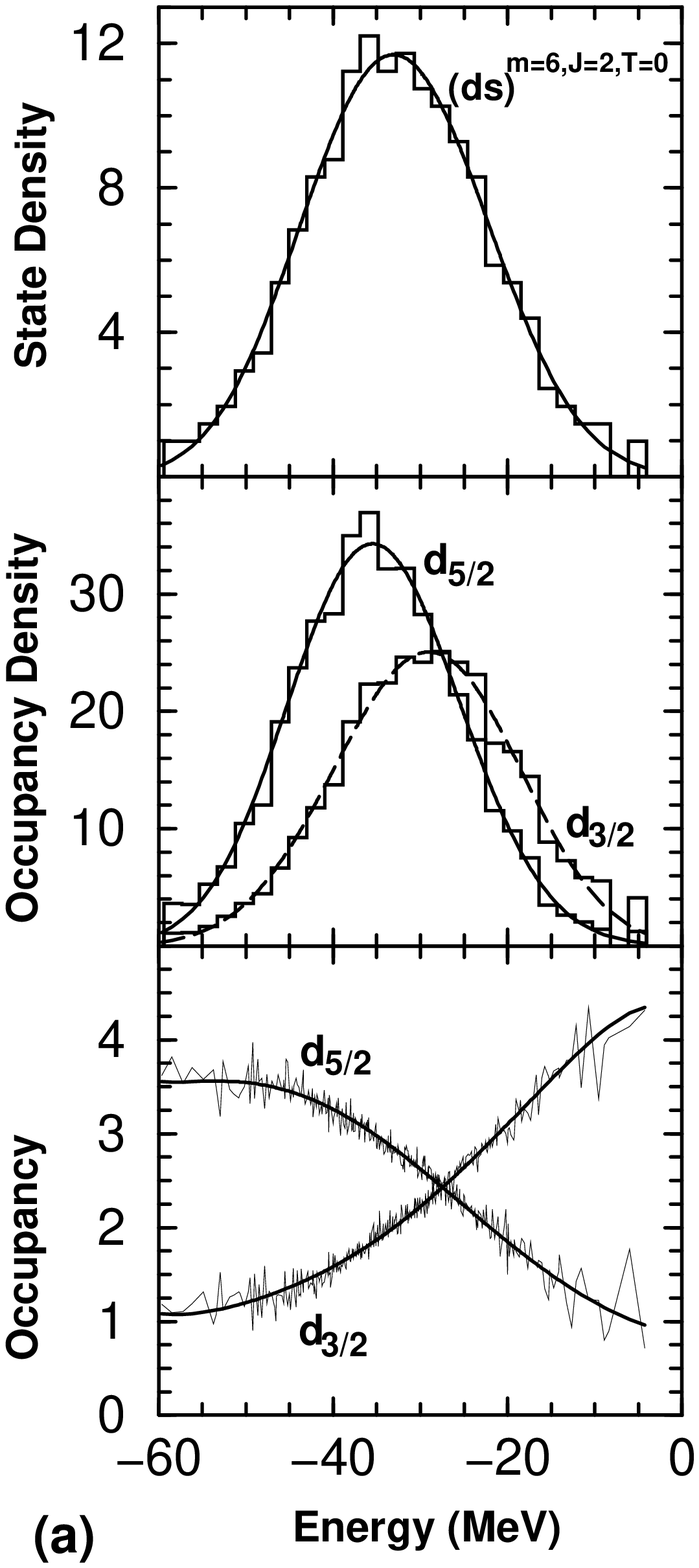}
&
\epsfxsize 2.5in
\epsfysize 6in
\epsfbox{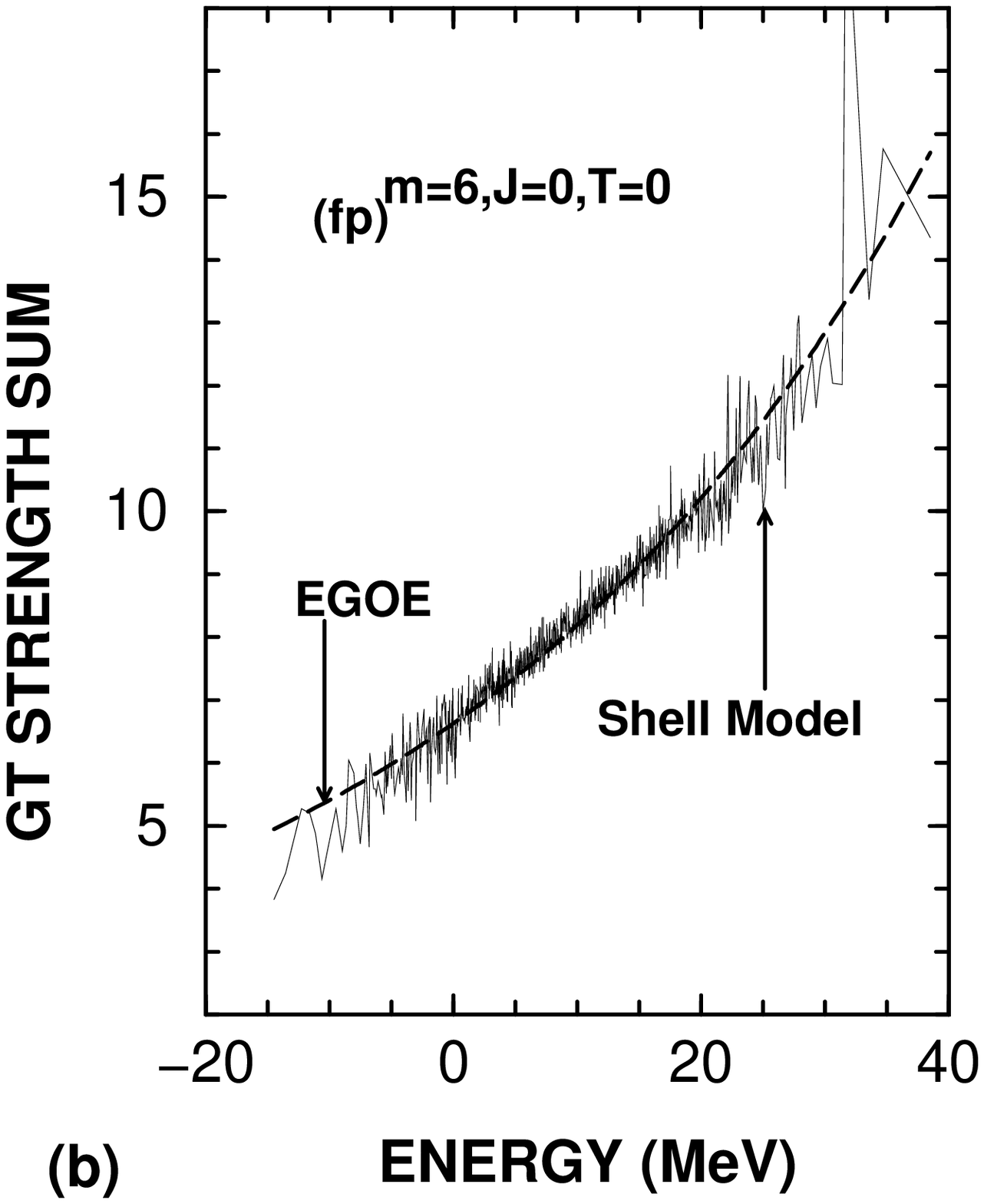}
\end{tabular}
\end{center}
\vskip -1cm
\noindent {\bf Fig. 2.} {\small {\bf (a)} Occupation numbers for $d_{5/2}$ and
$d_{3/2}$ orbits vs excitation energy ($E$) in the same example
as in Fig. 1b. Shown are also the shell model state and occupation number
densities (histograms) compared to the EGOE Gaussian forms
(continuous curves) given by (1,3).  {\bf (b)} GT strength sum versus
excitation energy ($E$) for the 814 dimensional six particle
$(fp)$-shell space with $J=0$, $T=0$.  The exact shell-model
results for the realistic KB3 interaction \protect\cite{Ca-95}
are compared with the EGOE predictions given by (3).}

\begin{center}
{\bf 3. Results with MF and SU(4) interpolating hamiltonians}
\end{center}

For further confirming the conclusions from Fig. 2, shell model calculations
are performed in the 325 dimensional 
$(ds)^{m=8,J=T=0}$ space  for occupancies and GT strength sums with
two different interpolating hamiltonians . First set of
calculations use the spherical shell model mean-field (MF) hamiltonian $h(1)$
as the unperturbed hamiltonian $H_0$ and in this case the occupation number
operators commute  with $H_0$,
\be
H_\lambda(MF) = h(1) + \lambda V(2) = H_0 +\lambda (H_{SM}-H_0)
\ee
Note that $H_{SM}=h(1)+V(2)$; the $h(1)$ is defined by $^{17}$O
single particle energies and $V(2)$ by Kuo's two-body matrix
elements as in the Sect. 2 $(ds)$-shell example.  In the figures
in Figs. 3a-d the calculations with (4) are denoted by {\bf MF}.
It is easily seen that spherical configurations (generated by
distributing the nucleons in the three $(ds)$ shell orbits) are
eigenstates for $H_0$. Therefore for $\lambda=0$ in (4), the
spectrum will have degeneracies.
In the second set of calculations the $SU(4)-ST$ scalar
part $H_{SU(4)-ST}$ of $H_{SM}$ is used as the unperturbed hamiltonian,
\be
H_\lambda(SU(4))=H_{SU(4)-ST:scalar} + \lambda (H_{SM}-H_{SU(4)-ST:scalar})=
H_0 +\lambda (H_{SM}-H_0)
\ee
In the figures in Figs. 3a-d calculations with (5) 
are denoted by {\bf SU(4)}.  Note that the GT
operator commutes with the SU(4) hamiltonian $H_{SU(4)-ST}$.
For $H=H_{SU(4)-ST}$, the eigenvalues and
eigenvectors are given easily by $SU(4)-ST$ algebra. The
eigenstates are labelled, for a given number $m$ of valence
nucleons, by $L$, $S$, $J$, $T$ and the $U(4)$ irreducible
representations (irreps) $\{f\}=\{f_1 f_2 f_3 f_4\}$ or the
$SU(4)$ irreps $\{F\}=\{F_1 F_2 F_3 \}$ where
$f_1+f_2+f_3+f_4=m$, $f_1 \ge f_2 \ge f_3 \ge f_4 \ge 0$,
$F_1=(f_1+f_2-f_3-f_4)/2$, $F_2=(f_1-f_2+f_3-f_4)/2$,
$F_3=(f_1-f_2-f_3+f_4)/2$ with additional restrictions on
$f_i$'s coming from the spatial part. In addition, as GT
operator is a generator of the $SU(4)$ group, the $SU(4)$
algebra gives directly $\lan K({\mbox{GT}}) \ran^E$ in terms of
$SU(4)-ST$ quantum numbers.
Methods for constructing $H_{SU(4)-ST}$ part of a given $H$ are
given in \cite{Ha-86},
\be
\barr{rcl}
H_{SU(4)-ST} & = & \frac{1}{2} (n-1)(n-2)\;E(0,\{0\} 0 0) - n(n-2)\;
E(1,\{1\} \frac{1}{2}  \frac{1}{2}) \nn8
& & + \dis\frac{1}{8} \l[n(2n-9) + G_2 +
2(S^2+T^2)\r]\;E(2,\{2\} 1 1) \nn8
& & - \dis\frac{1}{8} \l[n - G_2 +
2(S^2+T^2)\r]\;E(2,\{2\} 0 0) \nn8
& & + \dis\frac{1}{8} \l[n(n+3) - G_2 +
2(S^2-T^2)\r]\;E(2,\{11\} 1 0) \nn8
& & + \dis\frac{1}{8} \l[n(n+3) - G_2 -
2(S^2-T^2)\r]\;E(2,\{11\} 0 1) 
\earr
\ee
In (6) $n$ is number operator, $G_2$ is $U(4)$ quadratic
Casimir operator with eigenvalues $\lan G_2 \ran^{\{f\}} =
f_1(f_1+3) + f_2(f_2+1) + f_3(f_3-1) + f_4(f_4-3)$ and $S^2$ and
$T^2$ are operators with eigenvalues $S(S+1)$ and $T(T+1)$.
Construction of $H_{SU(4)-ST}$ requires the values for the
centriod energies (determined by the given $H$) $E(m,\{f\} S T)
= \lan H \ran^{m, \{f\} S T}$. For the $(ds)$-shell $H_{SM}$
hamiltonian (defined after (4)), they are $E(0,\{0\} 0 0)=0$, $E(1,\{1\}
\frac{1}{2} \frac{1}{2})= -2.302$MeV, $E(2,\{2\} 1
1)=-4.176$MeV, $E(2,\{2\} 0 0)=-2.975$MeV, $E(2,\{11\} 1
0)=-8.360$MeV and $E(2,\{11\} 0 1)=-7.048$MeV.  Using these,
$H_{\lambda}$ is constructed and then occupancies and 
$\lan K({\mbox{GT}}) \ran^E$ are calculated for various $\lambda$ values. 
The $SU(4)-ST$ algebra gives for $\lambda=0$,
$\lan K({\mbox{GT}}) \ran^E = \frac{2}{3} (\lan C_2(SU(4))
\ran^{\{F\}} -S(S+1))$; $\lan C_2(SU(4)) \ran^{\{F\}}$ = 
$F_1(F_1+4) + F_2(F_2+2) + F_3^2$.
In addition, the allowed $\{f\}S$
values for the $(ds)$-shell example are: $\{2222\}0$; $\{3221\}1$;
$\{3311\}2,0$; $\{332\}1$; $\{4211\}1$; $\{422\}2,0$; $\{431\}1,2,3$;
$\{44\}0,2,4$; $\{5111\}0$; $\{521\}1,2$; $\{53\}1,3$; $\{611\}1$; $\{62\}2$.
With these results, the GT curve for $\lambda=0$ case is
constructed as a check of the shell model calculations.  Each of
the $\{f\}S$ states will have several $L=S$ states and therefore the
states here have degeneracies. 

It is clearly seen from the results in Figs. 3a,b that the EGOE smooth form is
not a good approximation to the exact results in the case of regular motion.
For $\lambda \sim 0$ there are several (approximately) good quantum numbers
with nearby levels carrying different sets of quantum numbers and therefore 
expectation values show large fluctuations as a function of excitation energy.
The order-chaos transition as $\lambda$ increases is clearly illustrated by the
spectral regidity $\Delta_3$ in Figs. 3c,d  (also by the distribution of
nearest neighbour level spacings as shown in [14]).  It is seen that
occupancies and GT strength sums behave rather like the $\Delta_3$ statistic,
approching more slowly the EGOE and GOE limits respectively. This similarity is
probably due to the fact that both $\Delta_3$ and strength sums are related to
long-range correlations between the energy levels or wavefunctions.  Thus, in
the quantum chaotic domain transition strength sums, independent of the
hamiltonian, follow EGOE forms and we have statistical spectroscopy in the
chaotic domain. 

There is another important observation that follows from Figs. 3a-d, i.e. as
the interacting particle system becomes chaotic, expectation values take
smoothed forms (within GOE fluctuations \cite{Np-98}) and hence described by
the smoothed densities. For $\lambda >> \lambda_c$ ($\lambda_c$ corresponds to
order-chaos border and we determine $\lambda_c$ by using $\chi^2_{(\lambda)}$
which is the mean square deviation of the exact shell model strength sum from
the EGOE smoothed form \cite{Ko-99a}; see Fig. 4) they take the EGOE form given by (3). 
This generic result is of central interest in quantum chaos studies of finite
interacting particle systems as discussed in the next section. 

\begin{center}
{\bf 4. Further results and concluding remarks}
\end{center}

Besides the nuclear shell model results for GT strength sums and
occupation numbers presented in Figs. 2-4 (GT results for the
SU(4) case are reported in [14] recently), the behaviour of
occupancies as a many particle hamiltonian makes order-chaos
transitions is studied recently by several groups: (i)
using a 20 member EGOE(1+2) in 330 dimensional ${\cal N}=11,m=4$ space
with the MF hamiltonian (4); $h(1)$ is defined by the single particle
enegies $\epsilon_i=i+(1/i)\;;\;i=1,2,\ldots,11$ and $V(2)$ is EGOE(2)
\cite{Fl-97}; (ii) using the four interacting electrons Ce atom
\cite{Fl-94}; (iii) using a symmetrized
coupled two-rotor model \cite{Bo-98}; (iv) using EGOE(1+2) as
in (i) but in the 924 dimensional ${\cal N}=12,m=6$ space with 25
members \cite{Ko-99b}.  Most significant conclusion of
all these studies is that transition strength sums show quite
different behaviour in regular and chaotic domains of the
spectrum. In order to make this argument clear results of (iv)
are shown in Fig. 5; here occupancies for the single particle states
are calculated for various values of the interpolating parameter
$\lambda$.  It is clearly seen, for example from Fig. 5,
that below the region of onset of chaos  
\newpage

\begin{center}
\begin{tabular}{ll}
\epsfxsize 2.5in
\epsfysize 3.5in
\epsfbox{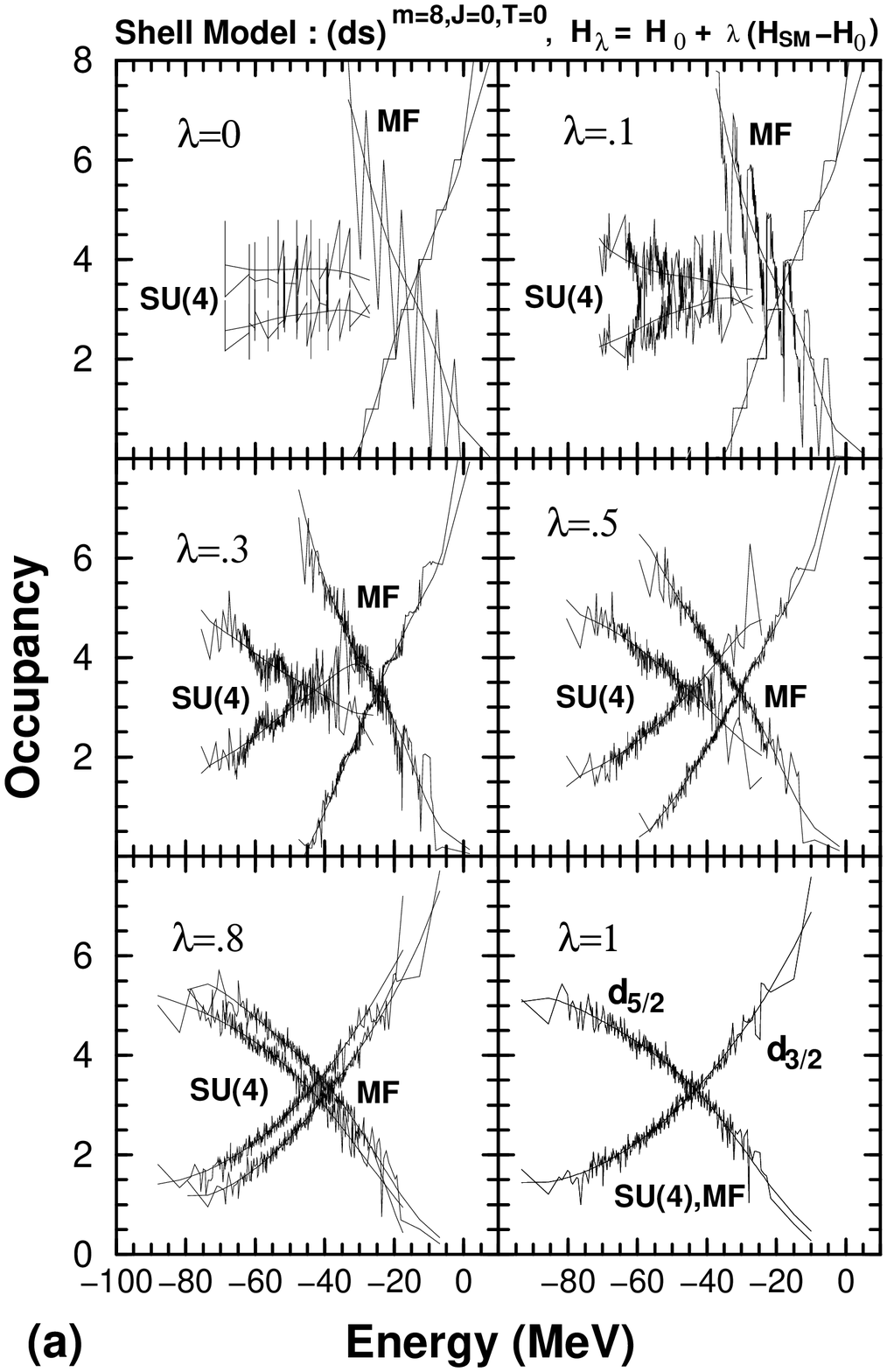}
&
\epsfxsize 2.5in
\epsfysize 3.5in
\epsfbox{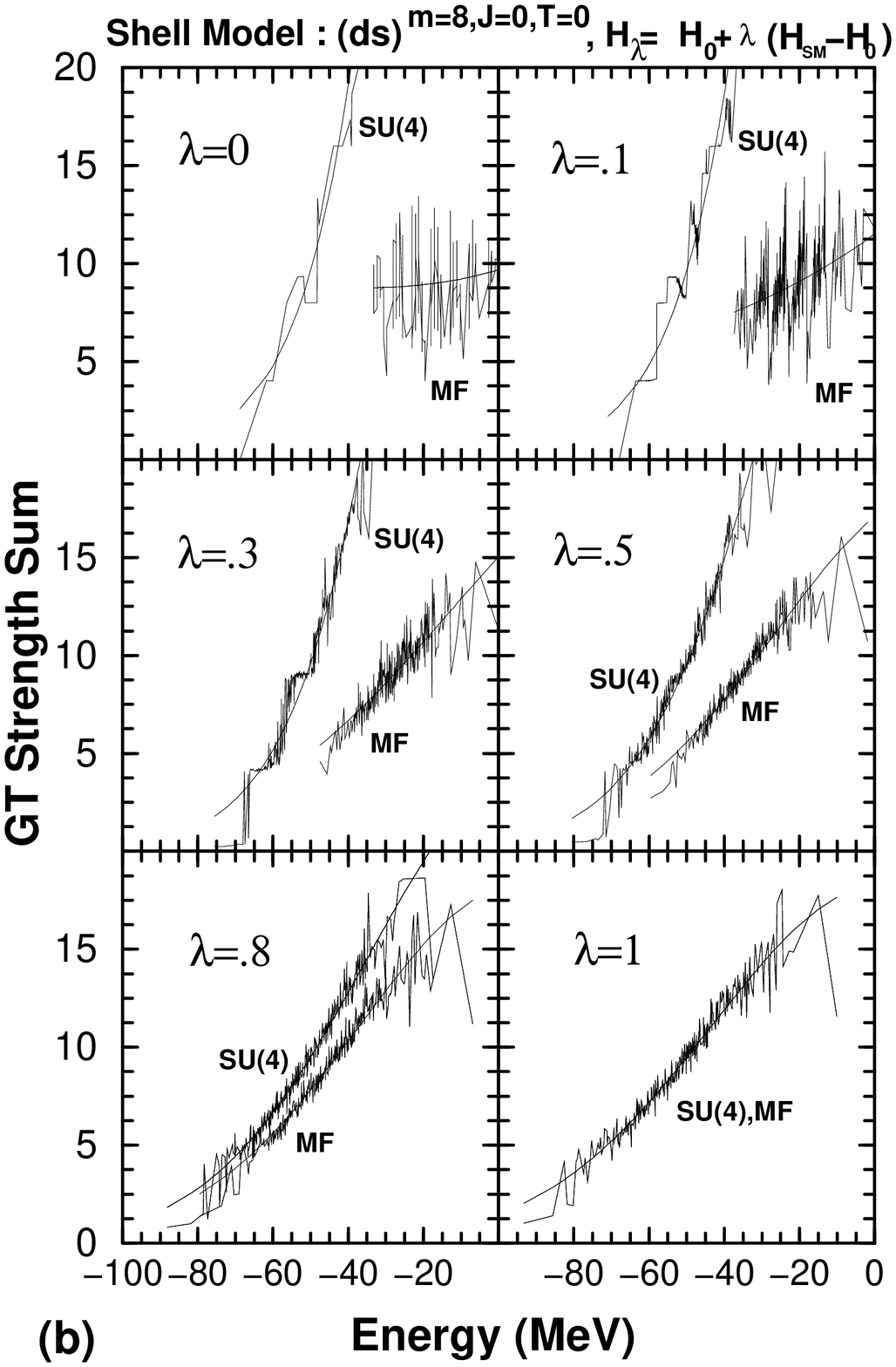} \\
\epsfxsize 2.5in
\epsfysize 3in
\epsfbox{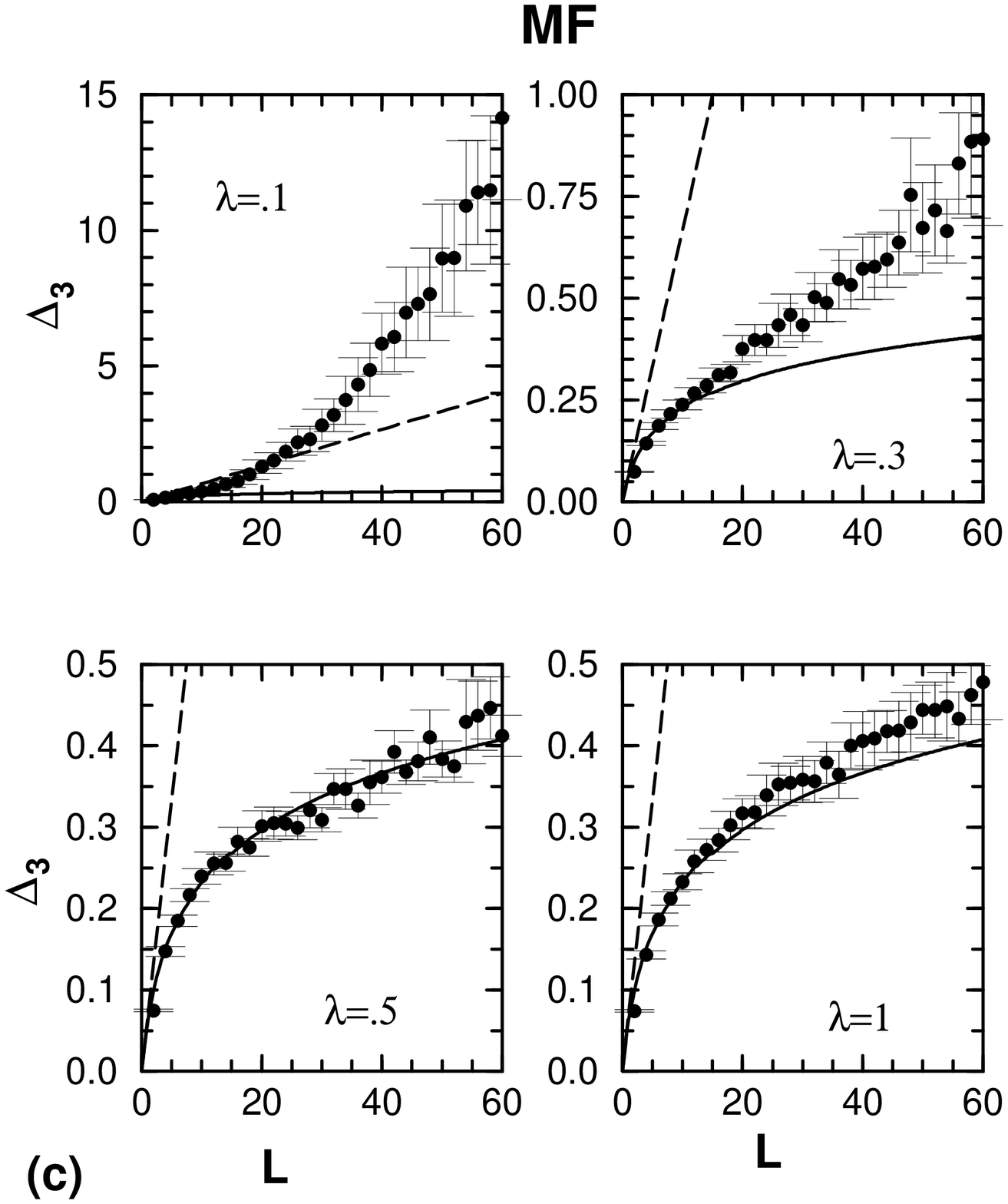}
&
\epsfxsize 2.5in
\epsfysize 3in
\epsfbox{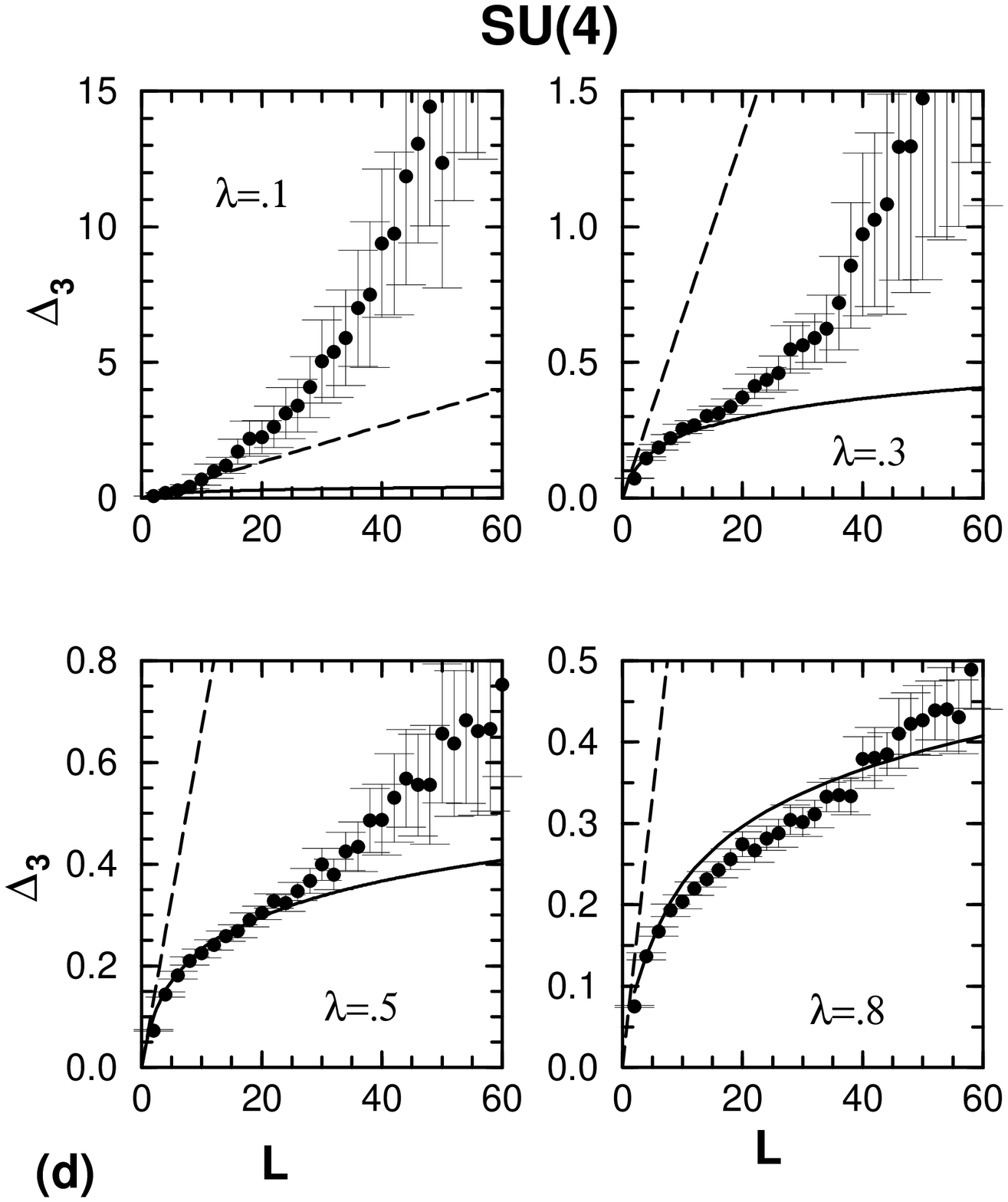} 
\end{tabular}
\end{center}
\vskip -1cm
\noindent {\bf Fig. 3.} {\small {\bf (a)} Occupation numbers for $d_{5/2}$ and
$d_{3/2}$ orbits vs excitation energy ($E$) for the MF and SU(4)
interpolating hamiltonians given by (4) and (5).  Shell model
results are compared with the EGOE predictions (continuous
curves) given by (3). {\bf (b)} same as (a) but for GT strength sums.
{\bf (c)} Averaged spectral rigidity $\Delta_3(L)$ for the eigenvalues
of the MF hamiltonian (4).  Error bars give the standard
deviation of the $\Delta_3$ average over overlaping intervals of
length $L$. The dashed curves are for Poisson and the continuous
curves are for GOE. {\bf (d)} same as (c) but for the SU(4)
hamiltonian (5). For the $\lambda=1$ case the parameters $(\epsilon, \sigma,
\gamma_1, \gamma_2)$ for the state, $d_{5/2}$, $d_{3/2}$ and GT densities are
($-52.59$ MeV, 13.15 MeV, 0.10, 0.03), ($-55.30$ MeV, 12.31 MeV, 0.01, $-0.06$),
($-48.70$ MeV, 13.42 MeV, 0.07, 0.05) and ($-48.37$ MeV, 12.64 MeV, 0.06, $-0.1$)
respectively.}

\newpage 
\begin{center} 
\epsfxsize 6in 
\epsfysize 7in 
\epsfbox{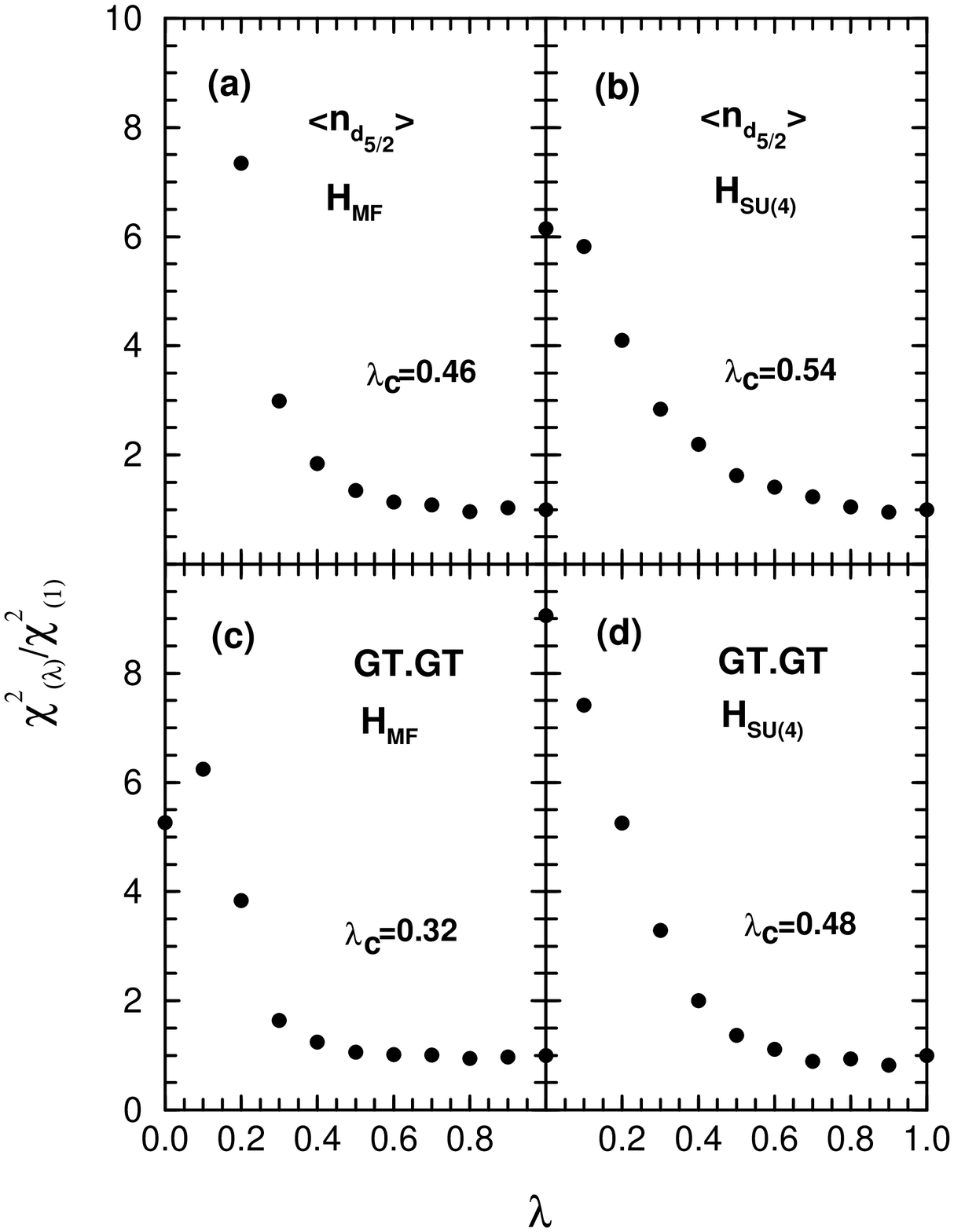}
\end{center} 
\vskip -2cm 

\noindent {\bf Fig. 4.} {\small $R(\lambda)=\chi^2_{(\lambda)}/\chi^2_{(1)}$ vs
$\lambda$. Results for MF and SU(4) calculations are shown for
$d_{5/2}$ occupancies and GT strength sums. From the observations in the
studies in \cite{Ko-99a}, a plausiable definition that the $H_\lambda$ system
is chaotic is given by the condition $R(\lambda) \leq 1.5$. Then, $\lambda_c$
is defined by $R(\lambda_c)=1.5$. The $\lambda_c$ values for the four cases
are shown in the
figure. The difference in the two $\lambda_c$ values in the case of occupancies
is easily understood in terms of the norms \cite{Fr-82,Kk-89} of $H_0$ and
$H_1=H_{SM}-H_0$; however this is not the case with GT strength sums. Thus it
appears that a complete theory for $\lambda_c$, in case of strength sums, may
come  from the study of $\Delta_3$, strength fluctuations, NPC and
information entropy for the hamiltonian and the transition operator involved.} 

\newpage
\begin{center}
\epsfxsize 6in
\epsfysize 7in
\epsfbox{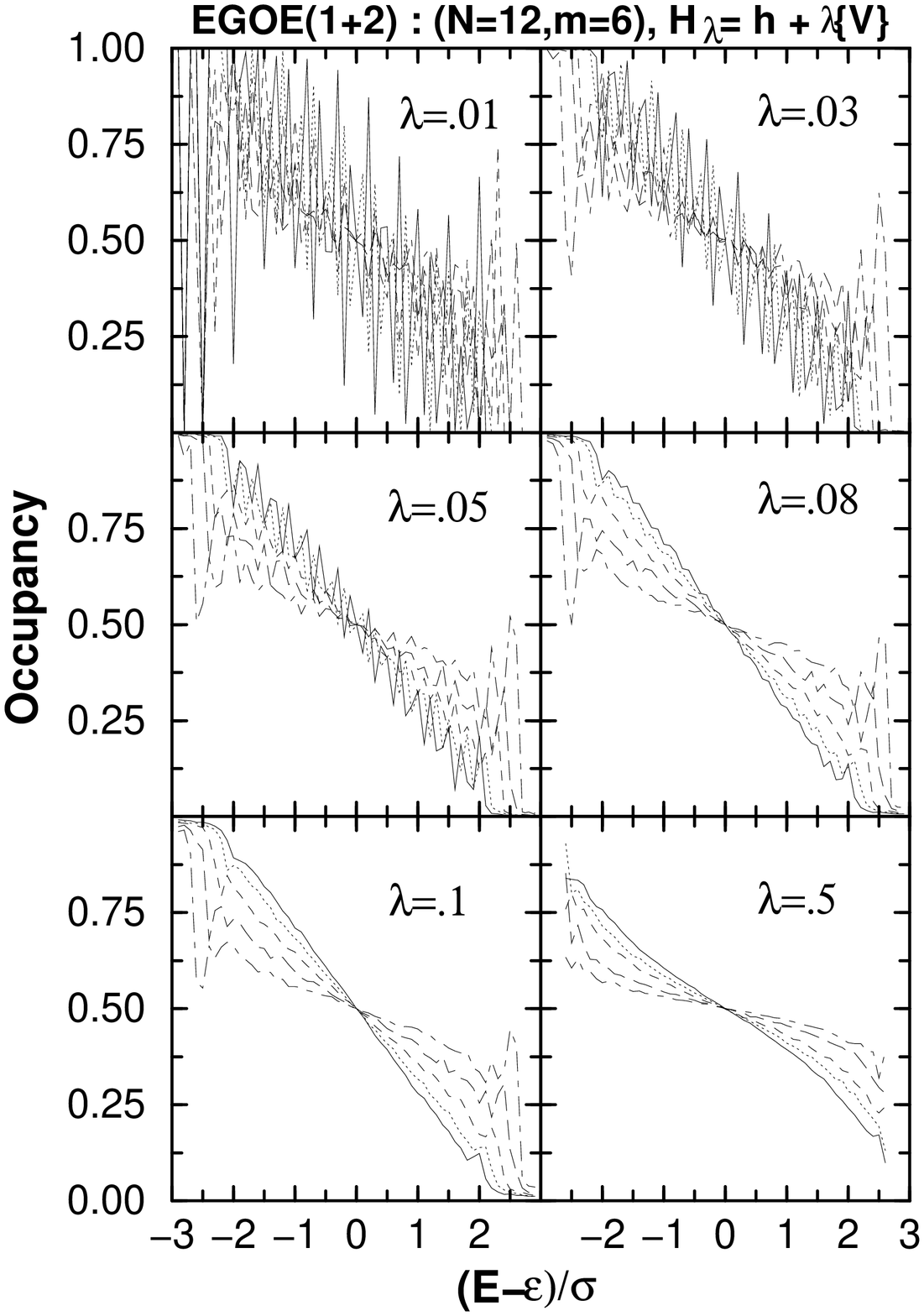}
\end{center}
\vskip -2cm
\noindent {\bf Fig. 5.} {\small Occupation numbers for a 25 member
EE(1+2) ensemble, defined by the hamiltonian
$h(1)+\lambda \{V(2)\}$, in the 924 dimensional ${\cal N}=12$, $m=6$
space (see text); in the figure ${\cal N}$ is denoted as $N$. Details of matrix
construction etc. are given in \cite{Ko-99b}.  Results are shown
for the lowest 5 single particle states and for six values of
$\lambda$. In the calculations occupation numbers are averaged
over a bin size of 0.1 in $(E-\epsilon)/\sigma$; $\epsilon$ is
centroid and $\sigma$ is width. The spectra of all the ensemble
members are first zero centered and scaled to unit width and
then the ensemble average is carried out. The estimate of
\protect\cite{Ja-97} gives $\lambda_c \sim  0.05$ for
order-chaos border in the present EE(1+2) example. It is clearly
seen that once chaos sets in, the occupation numbers take stable
smoothed forms. See \cite{Ko-99b} for further details.}

\newpage
\noindent occupation numbers show strong fluctuations (in the regular
ground state domain perturbation theory applies). In this region
there is no equilibrium distribution for $I_K(E)$, $I(E)$ and
other densities.  However in the chaotic domain (see
\cite{Fl-97,Ja-97} and also Fig. 4 for methods of determining $\lambda_c$) the
densities can be replaced by their smoothed forms as in (1)-(3).
Therefore there is a statistical mechanics, defined by various
smoothed densities, operating in the quantum
chaotic domain of finite interacting particle systems (this is
also the essence of statistical nuclear spectroscopy
[9,10,15-17]; see \cite{Fl-99} for arguments for 
statistical spectroscopy in atoms).
In fact in favourable situations, it is possible to introduce
thermodynamic concepts (effective temperatures and
chemical potencials etc.) in the chaotic domain. In order to firmly
establish the universality of these results, it is essential to
carry out numericl studies for a wide variety of interacting
particle systems and investigate various deformed EGOE's in detail.

\vskip 0.2cm
\noindent {\bf Acknowledgements}
\vskip 0.1cm
This work is partially supported by
DGES (Spain) Project No.  PB96-0604 and DST(India).

\ed